# High performing ionanofluid electrolyte with higher lithium salt concentration for safer lithium metal batteries

P. Bose, D. Deb and S. Bhattacharya

*Department of Physics; University of Kalyani, Kalyani; Nadia-741235; West Bengal. India.*

A new class of high-performance pyrrolidinium cation based ionanofluid electrolytes with higher lithium salt concentration are developed. The electrolytes are formed by dispersing imidazolium ionic liquid functionalized $TiO_2$ nanoparticles in low conducting, 0.6 M lithium salt doped N-alkyl-N-butylpyrrolidinium bis(trifluoromethylsulfonyl)imide ($Pyr_{14}TFSI$) ionic liquid (IL) hosted electrolyte. Viscosity, ionic conductivity and thermal properties of these electrolytes are compared with well-studied 0.2 M salt doped $Pyr_{14}TFSI$ IL-based electrolyte. The highly crystalline 0.6 M lithium salt dissolved IL-based electrolytes gradually become amorphous with the increasing dispersion of surface functionalized nanoparticles within it. The ionic conductivity of the electrolytes shows unusual viscosity decoupled characteristics and at the 5.0 wt% nanoparticle dispersion it attains a maximum value, higher than that of pure IL host. As compared to pure IL-based electrolytes, the ionanofluid electrolyte also possesses a significantly higher value of lithium ion transference number. The $Li/LiMn_2O4$ cell with the best conducting ionanofluid electrolyte delivers a discharge capacity of about 131 mAh $g^{-1}$ at 25 °C at a current density of 24 mA $g^{-1}$, much higher than that obtained in 0.2 M Li salt dissociated $Pyr_{14}TFSI$ electrolyte (87 mAh $g^{-1}$). Superior interfacial compatibility between ionanofluid electrolyte and electrodes as indicated by the excellent rate performance with outstanding capacity retention of the cell as compared to pure IL-based analogue, further establish great application potentiality of this optimized newly developed electrolyte for safer LMBs.

## Introduction

Lithium metal batteries (LMBs) containing lithium (Li) metal with highest theoretical capacity (3860 mAh$g^{-1}$) and the lowest negative potential (-3.04 V vs. standard hydrogen electrode) as the anode are reckoned as one of most efficient energy storage devices.[1] Nevertheless, the practical applications of secondary LMBs have not yet been initiated due to serious safety issues that originate from the use of conventional volatile and flammable aprotic carbonate-based electrolyte solvents.[2-4] These solvents have high reactivity with Li metal during repeated charge-discharge cycles, leading to uneven lithium deposition and lithium dendrite formation, which thereby reduce the stability of the cell and trigger short circuits and safety risks. [4] Therefore, serious efforts are being made worldwide to replace the carbonated electrolytes by nonvolatile and Li metal compatible electrolytes for the development of safer, high energy density secondary LMBs [5]. In this context, room temperature ionic liquid (IL) based electrolytes doped with lithium salts (LiX, with X being any anion) have shown immense potentials due to their brilliant physicochemical properties such as non-flammability, ultralow vapour pressure, excellent thermal stability and a wide range of electrochemical voltage stability. [5] Most of the IL-based electrolytes studied so far are either binary yLiX-(1-y)IL or ternary systems, containing a mixture of ionic liquids. [5-7] Among them quaternary ammonium-imide based N-alkyl-N-butylpyrrolidinium bis(trifluoromethylsufonyl)imide ($Pyr_{1A}TFSI$ ; A=3,4) ILs as solvents are found to be attractive due to their low cathodic limiting potential, and relatively better (>5 V) electrochemical stability window (vs. $Li/Li_+$ ) [8-9]. However, IL based electrolytes possess several undesired properties such as increased viscosity and thus lower ion conductivity with increasing Li salt dissociation. [6, 10, 11] But, on the other hand, LiX-IL electrolytes with low salt concentration suffer from low fraction of free electroactive $Li^+$ ions i.e. low lithium ion transference numbers ($t_{Li}^+$ < 0.2). Low $t_{Li}^+$ value often leads to a strong polarization effect in electrolytes owing to the formation of a concentration gradient during battery operation [11-14]. Although $t_{Li}^+$ may be increased by increasing salt concentration, a simultaneous significant decrease in conductivity reduces their battery performance. For example, xLiTFSI-(1-x) $Pyr_{14}TFSI$ electrolyte system with x=0.2 is found to be an attractive combination regarding ion conductivity and battery performance [15] as compared to its x=0.5 analogue. [16] Another important property that controls the performance of higher salt dissolved $Pyr_{14}TFSI$ based electrolyte is the higher crystallinity of the system. Thus, to incorporate higher ionic conductivity and moderately higher $t_{Li}^{+}$ in $Pyr_{14}TFSI$ IL-based electrolytes, higher salt content without crystallinity is essential. [14] Several strategies were implemented to improve the performance of aprotic pyrrolidinium IL-based electrolytes in LMB. These include its conversion from aprotic to protic, [16] incorporation of different additives such as dimethyl sulfite, [17] aprotic carbonates [18] within it, etc.

In this manuscript, we describe a strategy to develop a novel class of lithium ion conducting $PYR_{14}TFSI$-based electrolyte system for the application in LMBs. The electrolytes are prepared by dispersing different extents of $TiO_2$ nanoparticles tethered 1-n-butyl-(3-(trimethoxysilylpropyl) imidazolium bis(trifluoro- methanesulfonyl) imide nanoscale hybrid ionic fluids (NHIF), into the 0.6 M lithium salt doped $PYR_{14}TFSI$ IL based, low conducting, crystalline electrolyte. It has been observed that in the presence the NHIF the overall crystallinity of the system drastically subdued. Furthermore, despite a steady increase in macroscopic viscosity the ionic conductivity of system substantially improves and at 5.0 wt% NHIF the system exhibits an ionic conductivity higher than that of the 0.2 M salt doped electrolyte. Moreover, the lithium transference number of the newly designed ionanofluid electrolyte is substantially higher than that of the 0.2M electrolyte. The charge-discharge characteristics of LMBs assembled with the highest conducting electrolyte in combination with $Li/LiMn_2O_4$ electrodes are compared with that of the 0.2 M salt doped electrolyte containing similar system. Observed excellent rate capability and cycling performance with higher specific capacity of the ionanofluid electrolyte cell as compared to that of pure IL based system demonstrate the applicability of the electrolyte for safer LMBs.

## Experimental

### Synthesis

Anatase titania ($TiO_2$) nanoparticles were synthesized according to our earlier report. [19] The surface modification of as-synthesized $TiO_2$ nanoparticles was done by silane grafting reaction in two steps. Initially, equimolar 1-Butylimidazole (98.0%, Sigma Aldrich) and (3-Chloropropyl) trimethoxysilane (97%, Sigma Aldrich) were dissolved in dimethylformamide (DMF) and refluxed at 120 °C under a nitrogen atmosphere for 2 days. Resultant orange viscous honey-like hydrophilic 1-*n*-butyl-(3-(trimethoxysilyl-propyl) imidazolium chloride IL was purified via liquid extraction in ether followed by solvent evaporation chromatography. $^{1}H$ NMR (Fig. S1a)† (400 MHz, DMSO-$d_6$): δ (ppm) = 10.796 (s, 1 H), 7.568 (t, $J$ = 1.8 Hz, 1 H), 7.45 (t, $J$ = 1.2 Hz, 1 H), 4.334 (dt, $J$ = 4.2 Hz, 4 H), 3.571 (s, 9 H), 1.872–2.075 (m, 4 H), 1.352 (dt, $J$ = 8.1 Hz, 2 H), 0.967 (t, $J$ = 7.5 Hz, 3 H), 0.616–0.672 (m, 2H).

Excess amount (1.5 times) of as-prepared IL was allowed to react with 1.0 wt.% deionised water dispersed $TiO_2$ nanoparticles under an $N_2$ atmosphere for 12 h under continuous stirring at 100 °C, and subsequently, ethanol was added to the mixture. After cooling, $TiO_2$ nanoparticles grafted hydrophilic hybrid ionic fluids were collected by repeated washing. Finally, the hydrophilic ionic liquid converted to hydrophobic by metathesis reaction using aqueous solution of Lithium bis(trifluoromethylsulfonyl) imide (LiTFSI) salt (98%, TCI). The as-prepared $TiO_2$ core tethered nanoscale hybrid ionic fluid ($TiO_2$-NHIF) was washed several times with DI water and acetone to remove any unreacted elements and dried under dynamic vacuum at 100 °C for 24 h and stored in an Argon filled dry box for further use.

The hydrophilic ionic liquid 1-n-Butyl-1-methyl pyrrolidinium bromide ($PYR_{14}Br$)

was synthesized by refluxing equimolar 1-Methylpyrrolidine (98.0%, TCI) and 1-Bromobutane (99%, Sigma Aldrich) in acetonitrile at 80 °C under nitrogen atmosphere for 2 days, followed by purification by liquid extraction in ether. $^1H$ NMR (Fig. S1b)† (400 MHz, $CDCl_3$): δ (ppm) = 3.817-3.781 (m, 4 H), 3.669-3.626 (m, 2 H), 3.273 (s, 3 H), 2.295 (br s, 4 H), 1.708–1.788 (m, 2 H), 1.399–1.454 (m, 2 H), 0.984-0.947 (t, 3H, $J$ = 7.4 Hz). Similar metathesis reaction using aqueous solution of LiTFSi salt was performed to prepare hydrophobic $Pyr_{14}TFSI$ host.

Two $Pyr_{14}TFSI$ IL-based electrolytes were prepared by dissolving predetermined quantities of Lithium bis(trifluoromethane sulfonyl) imide (LiTFSI) and lithium bromide (LiBr) salts such that the total lithium salt concentration in the electrolyte was kept at either x=0.2 M or 0.6 M (i.e. xM(0.7LiTFSI+0.3LiBr) in $Pyr_{14}TFSI$). 30.0 wt% of lithium bromide (LiBr) was added with LiTFSI as dopant salt to improve the cycle life of LMBs by suppressing the dendrite formation or short-circuiting, as was reported earlier. [20] The resultant electrolytes were dried under dynamic vacuum at 80 $^0$C for two days and kept in an argon-filled dry box for further use. The electrolytes are represented as $xLiPyr_{14}TFSI$ (x=0.2, 0.6) in the manuscript.

The ionanofluid electrolytes were obtained by incorporating y wt% (y=1.0, 2.5, 5.0, 7,5 and 10.0) of $TiO_2$-NHIF in the $0.6LiPyr_{14}TFSI$ electrolyte in an argon atmosphere, under slow stirring. The final composition of the ionanofluid electrolytes became yNHIF-(100-y) [(0.6(0.7LiTFSI +0.3LiBr) in $Pyr_{14}TFSI$] and are denoted as yNHIF0.6LiPy in the manuscript.

## Analysis methods

The chemical structures of the ionic liquids and NHIF were determined by $^1H$ NMR on a Bruker Avance-400 MHz NMR Spectrometer. Transmission electron microscopy (TEM) of the hybrid ionic fluid and different ionanofluids were performed on a JEOL JEM-2100, 200 kV microscope in high-resolution mode.

Thermogravimetric (TGA) scans were performed within a temperature range from room temperature to 700 $^0$C using a Netzsch TG 209 F3 Tarsus system at a heating rate of 10 °C/min with equal sample mass (15±0.5) mg (Fig. S2)†. Calorimetric measurements were carried out with a Netzsch DSC 214, Polyma differential scanning calorimeter (DSC) under nitrogen atmosphere. Before measurement both temperature and enthalpy were auto-calibrated by pure indium. In a typical experiment ~15 ml sample was preheated to 120 °C @ 10 °C/min and kept isothermally for 10 min to eliminate any thermal history. After that, the sample was rapidly quenched (@ 80 °C/min) to -100 °C, followed by an isothermal step for 10 min. Finally, the supercooled liquid was allowed for cold crystallization through heating up to 30 °C. Viscosity measurement of different ionanofluid electrolytes was carried out in the temperature range of 25 to 100 °C in steps of 5 °C, using an advanced air bearing rheometer, model Kinexus pro$^+$ (Malvern Instruments, UK) attached with cone and plate geometry.

The ionic conductivity as a function of frequency (Fig. S4)† was evaluated by impedance spectroscopy measurements over the frequency range of 42 Hz to 8 MHz with an amplitude of 5 mV in the temperature range from -20 °C to 100 °C using a Hioki LCR meter model IM3536 under dynamic vacuum. The temperature was controlled by Eurotherm temperature controller with temperature constancy of ±0.1 $^0$C. Electrochemical properties of the electrolytes were measured using electrochemical workstation (CHI608E).

To evaluate the lithium transference numbers ($t_{Li}^+$) of the ionanofluid electrolytes, Symmetric lithium coin cells (type 2032) (Li|electrolyte|Li) were assembled in an argon filled glovebox (MBraun. Labmaster), with a glass fibre filter (Whatman, GF/A) as the separator. Lithium metal foil with a thickness of 0.75 mm was procured from Alfa Aesar (99.9%, metals basis) and used as received. The $t_{Li}^+$ was measured by chronoamperometry and electrochemical impedance spectroscopy at room temperature using a multichannel electrochemical analyser analyser (Bio-Logic, model VMP3).

Lithium manganese oxide, $LiMn_2O_4$ (LMO) with two-dimensional layered crystal structure are of great interest as cathode materials for rechargeable lithium-metal batteries. [21] This electrode is found useful for long cycle life, high power applications, safety, low cost, and low toxicity. [22] In this study, LMO electrode was fabricated from a mixture of LMO powder (MTI corporation, USA), acetylene black and PVDF in N-methyl-2-pyrrolidone solvent with a weight ratio of 80: 10: 10. The loading of the active material was ca. 2.5–3.0 mg/cm$^2$. CR2032-type coin cells were assembled by compacting in sequence a LMO electrode, a Whatman glass fibre filter (GF/A) separator soaked with electrolytes and a lithium foil in an argon filled glovebox (MBraun. Labmaster). The charge-discharge properties (in terms of capacity, cyclic stability, and rate capability) of different cells were compared using a Netware battery testing system within a voltage range of 2.8–4.4 V and at different current densities.

## Results and discussion

Fig. 1(a-b) depict the bright field TEM micrographs of $TiO_2$-NHIF at different magnifications. Uniform dispersion of individual nanocrystals with an average diameter of 12-14 nm (Fig. 1(b)) can be observed. Selected area energy diffraction (SAED) obtained from the group of nanoparticles (Fig. 1(c)), shows ring patterns corresponding to the diffraction from (101), (004), (200), (211), and (204) crystal planes of tetragonal anatase $TiO_2$. The HRTEM images (Fig. 1(d)) display lattice fringes with a d-spacing of 0.35 nm, corresponding to the (101) plane of anatase $TiO_2$, corroborating the SAED patterns (JCPDS No. 00-021-1272). Fig. 1(e) displays the dispersion of 5.0 wt% $TiO_2$-NHIF in the $Pyr_{14}$TFSI IL host. As can be observed, the surface functionalized $TiO_2$ nanoparticles are uniformly spread over the host IL to form stable ionanofluids.

Thermal stability of different electrolytes is compared with pure IL in Fig. S2†.

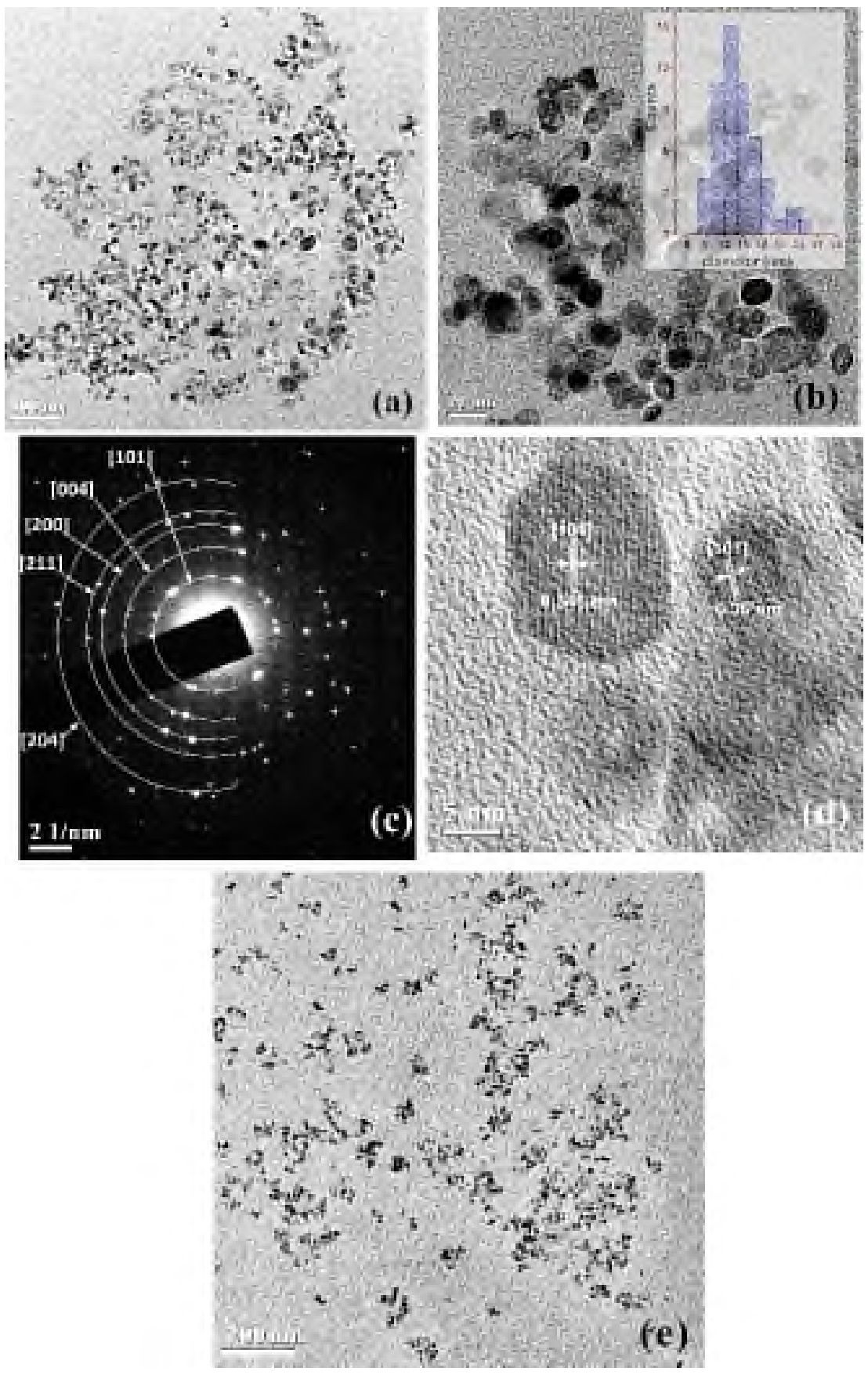

**Fig. 1** (a-b) TEM images of the $TiO_2$-NHIF at different magnifications. Size distribution of the nanoparticles are shown at the inset of (b). (c) SAED patterns showing ring patterns corresponding to the diffraction from $TiO_2$ crystal planes (d) High-resolution image showing lattice spacing. (e) Dispersion of 5.0 wt% NHIF in $Pyr_{14}$TFSI ionic liquid.

As shown, both pure IL and the electrolytes are thermally stable up to ~400 °C and the presence of either salt concentration or NHIF have minimum influence on it.

Fig. 3 illustrates the DSC traces for different samples obtained during the heating scans @5 °C/min from their respective supercooled states. Above the glass transition temperature (Tg=-86.5 °C), the endotherm of $Pyr_{14}$TFSI IL displays the evidence of metastable phase (melts at -27.7 °C) along with sharp undercooled liquid to solid phase transition at -53.7 °C, (melts at higher temperature -17.8 °C). This is in agreement with the previous study. [23] For the 0.2 M lithium salts doped electrolyte, $T_g$ shifts to higher temperature (-85 °C) (Fig. S2) with a sharp reduction in overall crystallinity of the system (Table S1, supporting information). However, with the addition of higher salt concentration (0.6 M) the system again become highly crystalline, exhibiting polymorphism with a large shift in melting endotherms (-6.5 °C).This is in line with the earlier reports [24, 14] that higher concentration of lithium salt promotes the crystallization of the electrolytes. For the ionanofluid electrolytes formed by the dispersion of NHIF in the 0.6LiPyr$_{14}$TFSI electrolyte, the melting temperature again shifts to a lower temperature, close to that of the pure IL. With increasing NHIF amount, both $T_g$ and crystallization temperature of the system gradually shift to higher temperature. Furthermore, both crystallization and melting enthalpy of the ionanofluid electrolytes show a decreasing trend with increasing amount of NHIF (Table S1)†, indicating a suppression of crystallinity. Consequently, the endotherm of 10Li0.6Py ionanofluid does not show any phase transition, exhibiting only $T_g$ at nearly -77.5 °C (Fig. S2)†. Similar suppression of supercooled to crystalline phase transition was also observed earlier for $Al_2O_3$-NHIF dispersed imidazolium-based ionanofluid system [25]. Fig. 4(a-b) illustrate the temperature dependent steady-state viscosities and conductivities (obtained from the analysis of the real part of conductivity spectra (Fig. S3†) of different electrolytes along with pure Pyr$_{14}$TFSI IL The pure IL is characterized by the lowest viscosity with a consequent high value of ionic conductivity. For the 0.2Pyr$_{14}$TFSI electrolytes, observed moderate increase in viscosity corresponds to a slight decrease in conductivity, ascribing to the small contribution of the $Li^+$ or $TFSI^-$ ions to the total number of ions in the system. However, for the electrolyte with 0.6 M salt concentration, a considerable increase in viscosity with observed higher crystallinity (Fig. 2) lead to a sharp decrease in conductivity. The observed feature is in line to the classical Walden law, suggesting a viscosity controlled ion conduction in these electrolytes. However, it is interesting to observe that with the incorporation of NHIF in the electrolyte, despite a further increase in viscosity the ionic conductivity of the system remarkably improves.

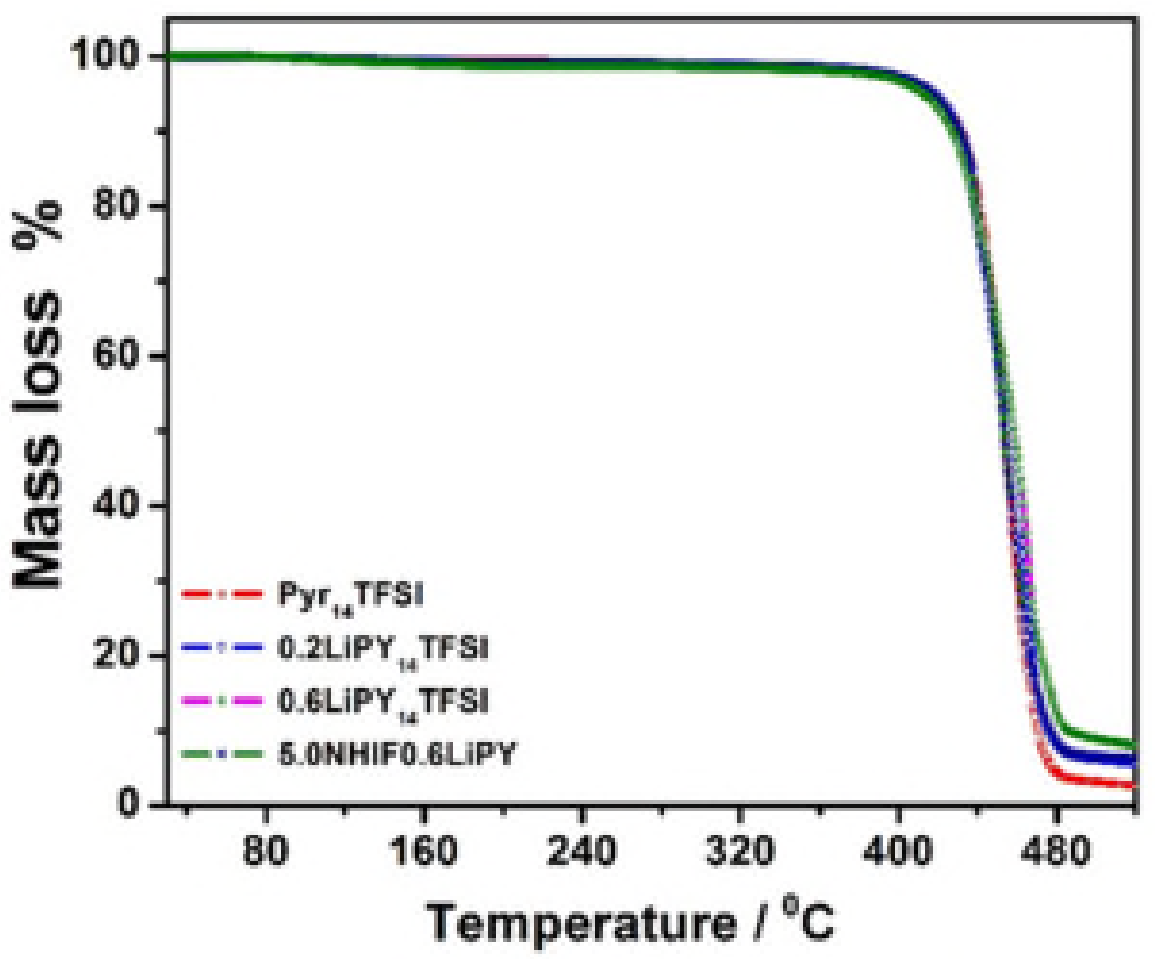

**Fig 2** Comparison of thermal stability of different electrolytes with Pyr$_{14}$TFSI ionic liquid.

This is in contrast to the Walden law and the unusual phenomena continue up to 5.0 wt% NHIF incorporation and for the as-prepared 5.0NHIF0.6LiPy ionanofluid electrolyte, the ionic conductivity attains a maximum value of 3.0 mS cm$^{-1}$ at303 K.

The ionic conductivity falls sharply with further increase of NHIF dispersion in the electrolyte. Observed non-Arrhenius temperature dependence of both the viscosity and ionic conductivity can be suitably described by the Vogel–Fulcher–Tammann (VFT) equations [26] expressed as

$$\eta = \eta_0 \exp(E_{a\eta} / k_B(T - T_0)) \qquad (1)$$

$$\sigma_{dc} = \sigma_0 \exp(-E_{a\sigma} / k_B(T - T_0)) \qquad (2)$$

Here, $E_{a\eta}$ and $E_{a\sigma}$ are the activation energies for viscous flow and ion conduction, respectively. $T_0$ is considered as the equilibrium glass transition temperature where molecular mobility vanishes i.e. flow resistance becomes maximum. Its value is usually considered as about 30 °C lower than the thermodynamic glass transition temperature ($T_g$). Solid lines in Fig. 3 are the best fits to the VFT expressions in Eq. (1) and (2), with the fitting parameters as shown in Table 1. Identical value of $T_0$ has been used for both the analysis.

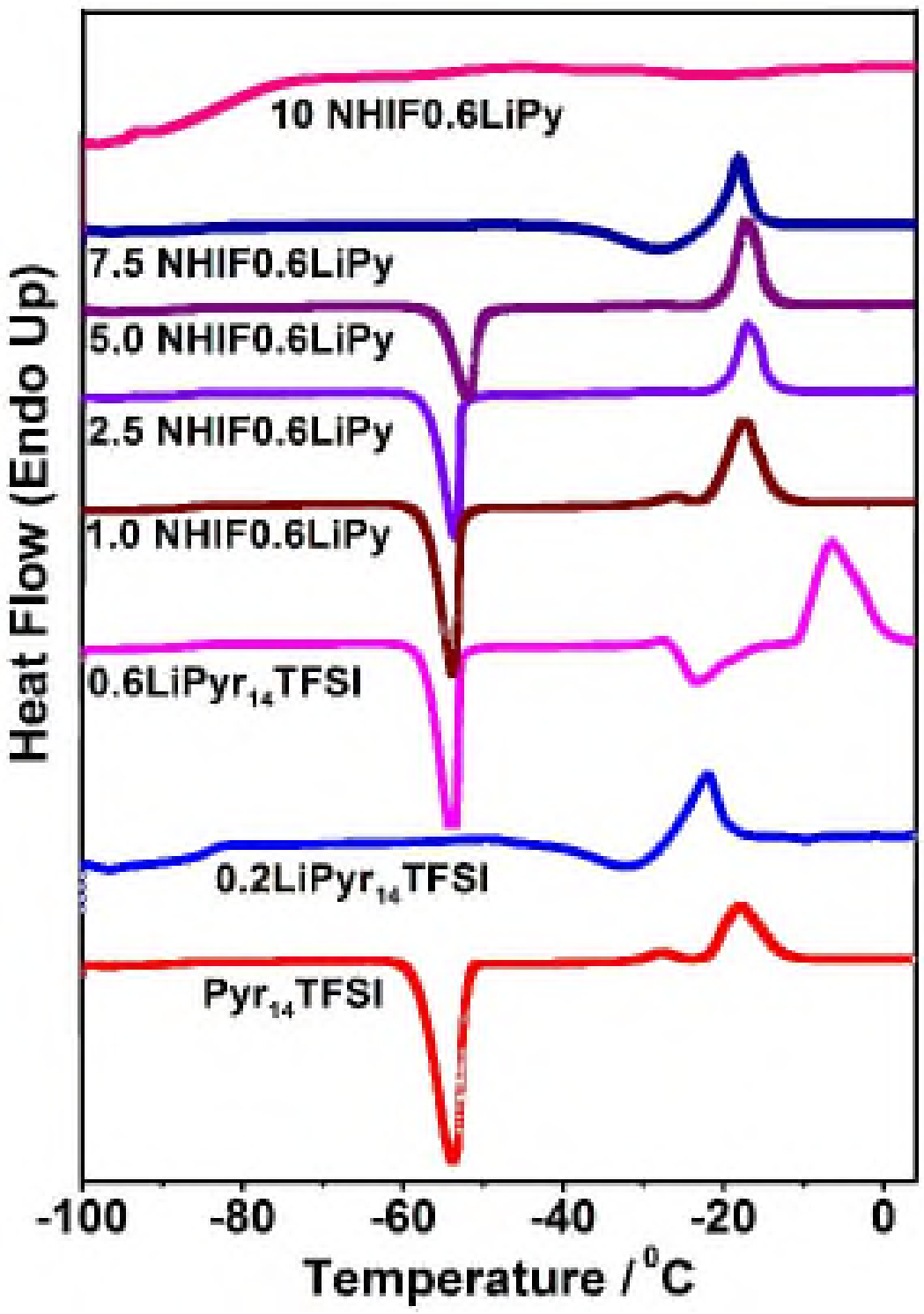

**Fig. 3** DSC thermograms of pure $Pyr_{14}TFSI$ IL and different electrolytes during heating (φ=5 °C/min) from their respective supercooled states.

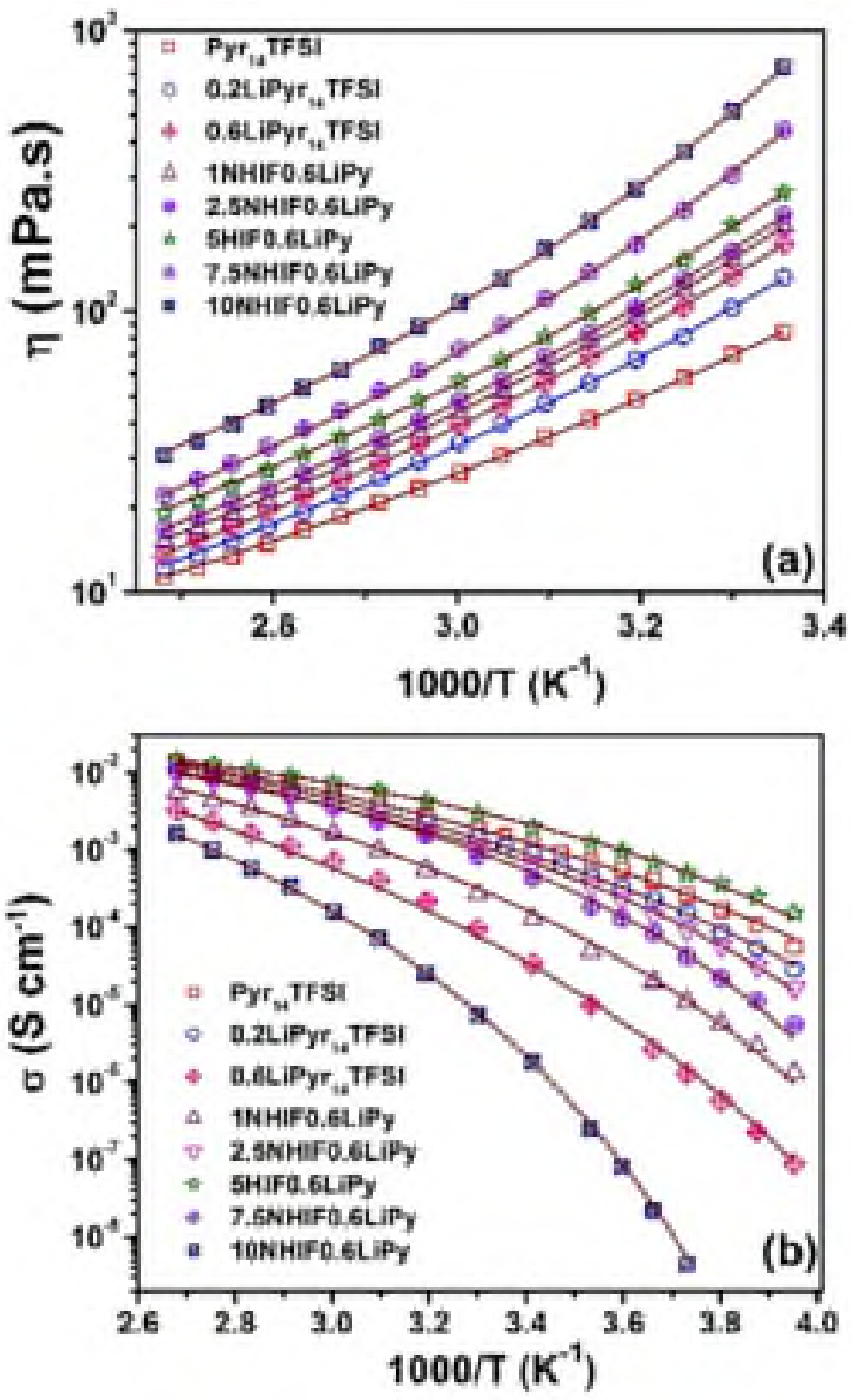

**Fig. 4** Arrhenius temperature variation of (a) viscosity and (b) Dc conductivity for different electrolytes along with pure $Pyr_{14}TFSI$ IL.

As observed from Table 1, both $E_{a\eta}$ and $T_0$ gradually increase with increasing concentration of either salt or NHIF or both, within the host IL. It is observed that for exclusive salt dissociated electrolytes, the activation energy for ion conduction $E_{a\sigma}$ increases with increasing salt concentration. However $E_{a\sigma}$ progressively decreases with increasing extent of NHIF and attains minima for the 5NHIF0.6LiPy electrolyte, above which it increases drastically. In this concern it is worthy to mention that according to the earlier reports, [12-14] the addition of lithium salts in ILs attracts the loosely bound anions of the IL towards the relatively smaller $Li^+$ ions, leaving the larger organic cations free.

Higher salt concentration further enhances the ion-pairs and a redistribution of charge species occurs in the system. Consequently, a slowdown in the overall ion dynamics i.e. a substantial reduction in the $Li^+$ ion mobility occur [14]. Observed sharp decrease in ionic conductivity with higher activation energy ($E_{a\sigma}$) for conduction, despite higher $Li^+$ ion density, indicate the same. Higher crystallinity of the system further reduces its conductivity. In our earlier report, it has already been established that nanoparticle tethered NHIF is very effective for reducing ion pair formation in ionanofluid [25]. Likewise, in the present case, $TiO_2$-NHIF gradually weakens the strong bonding between the $Li^+$ cations and IL anions, thereby releasing more and more $Li^+$ ions to contribute to the ion conduction. That results in a significant increase in conductivity. A gradual reduction in overall crystallinity also favours the system overcoming the detrimental effect of the simultaneous increase in viscosity.

A steady decrease in $E_{a\sigma}$ along with a corresponding increase in $E_{a\eta}$ also supports the same. However, for the electrolyte with NHIF content above 5.0 wt%, although the crystallinity of the system substantially reduces, the sufficiently higher flow resistance supersedes all the positive effects of NHIF, resulting in a sharp decrease in conductivity with a higher value of $E_{a\sigma}$ .

**Table 1** Compositional variation of VTF parameters obtained from the fitting of temperature dependent viscosity and dc conductivity data.

| Samples | Viscosity | | | Dc conductivity | | |
|---|---|---|---|---|---|---|
| | $\eta_\infty$ [mP s] | $E_{a\eta}$ [×10$^{-2}$ eV] ±2.0×10$^{-3}$ | $T_0$ [K] | $\sigma_\infty$ [S cm$^{-1}$] | $E_{a\sigma}$ [10$^{-2}$ eV] ±2.0×10$^{-3}$ | $T_0$ [K] |
| $Pyr_{14}TFSI$ | 0.28 ±0.02 | 6.7 | 160.5 | 0.77±0.02 | 7.0 | 160.5 |
| $0.2LiPyr_{14}TFSI$ | 0.23 ±0.03 | 7.0 | 168.2 | 0.75±0.02 | 7.6 | 168.2 |
| $0.6LiPyr_{14}TFSI$ | 0.32 ±0.01 | 7.4 | 175.0 | 0.18±0.01 | 11.1 | 175.0 |
| 1NHIF0.6LiPy | 0.34 ±0.01 | 6.3 | 183.8 | 0.23±0.03 | 8.2 | 183.8 |
| 2.5NHIF0.6LiPy | 0.42 ±0.02 | 5.9 | 188.4 | 0.30±0.02 | 5.5 | 188.4 |
| 5NHIF0.6LiPy | 0.51 ±0.02 | 5.4 | 190.2 | 0.51±0.03 | 4.1 | 190.2 |
| 7.5NHIF0.6LiPy | 0.45 ±0.02 | 5.8 | 199.7 | 0.43±0.02 | 5.2 | 199.7 |
| 10NHIF0.6LiPy | 0.73 ±0.01 | 7.6 | 208.2 | 1.15±0.02 | 8.9 | 208.2 |

Ionic conductivity in IL-based lithium electrolytes resembles the mobility of three different ions (i.e. IL and $Li^+$ cations, and anions) in the system [14]. Among these charge species, the transport number of $Li^+$ ($t_{Li}^+$) is the most important as it has a direct consequence to the reduction in polarization in electrolyte and improvement of power density in LMBs. From the measurements of potentiostatic DC polarization (chronoamperometry) under an applied voltage $\Delta V$=10 mV, in combination with electrochemical impedance spectroscopy, $t_{Li}^+$ for an electrolyte can be evaluated following the equation proposed by Bruce and co-workers [27] as

$$t_{Li^+} = I^S(\Delta V - I^0 R_I^{\,0}) / I^0(\Delta V - I^S R_I^{\,S}) \qquad (3)$$

where $I^s$ and $I^0$ are the initial (polarized state) and steady-state (unpolarized state) currents, respectively. Similarly, $R_I^{\,0}$ and $R_I^{\,S}$ are the initial and steady-state resistances of the passivation layers, respectively. Li-ion transference numbers, ($t_{Li}^+$) of the low (0.2 M) and high (0.6 M) salt doped electrolytes are compared with the ionanofluid electrolyte with highest ion conductivity (5NHIF0.6LiPy). Fig. 4(a-b) display the Nyquist plots of impedance before and after the chronoamperometry measurements of two Li-symmetrical cells, comprising $0.2LiPyr_{14}TFSI$ and 5NHIF0.6LiPy electrolytes, at room temperature (303 K). Inset of the figures show the corresponding DC polarization currents. Using Eq. (3) and from the experimental data, the value of $t_{Li}^+$ for the electrolytes are evaluated as 0.24 and 0.86, respectively. Similarly $t_{Li}^+$ for lowest conducting $0.6LiPyr_{14}TFSI$ electrolyte is estimated as 0.64.

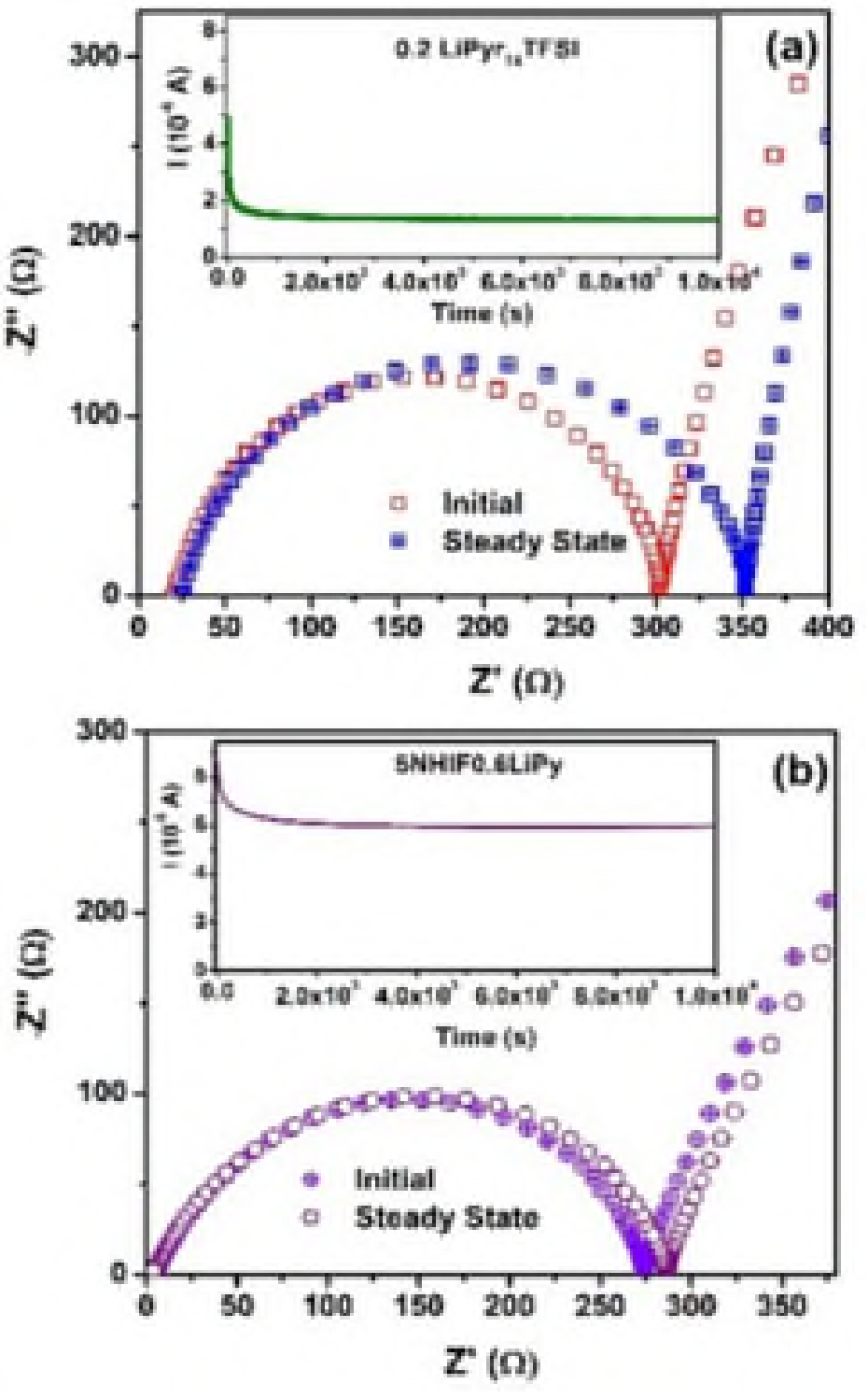

**Fig. 5** Nyquist Plots of initial and steady-state impedance for (a) $0.2LiPyr_{14}TFSI$ and (b) 5NHIF0.6LiPy electrolytes. Respective chronoamperometry profiles are shown at insets.

The value of $t_{Li}{}^{+}$ obtained for 0.2 M salt doped electrolyte is significantly higher than that estimated ($t_{Li}{}^{+} \approx 0.13$) from the measurement of individual self-diffusion coefficients of the anions and cations using PFG-NMR [12, 14]. To accommodate the contribution of diffusion in the unpolarized and polarized states for symmetric electrolytes, Lu et. al. had proposed a modification in the Bruce equation as [28]

$$t_{Li^+} = I^S(\Delta V - I^0 R_I{}^0)/2I^0(\Delta V - I^S R_I{}^S) \qquad (4)$$

It is interesting to observe that the value $t_{Li}{}^{+}$ for 0.2LiPy$_{14}$TFSI electrolyte estimated using Eq. (4) is 0.12, which is close to the value obtained from PFG-NMR [12, 14]. Accordingly, the values of 0.6LiPy$_{14}$TFSI and 5NHIF0.6LiPy electrolytes are found to be 0.32 and 0.43 respectively. This is plausible as the higher salt concentration improves the $t_{Li}{}^{+}$ and release of more $Li^+$ ions in the electrolyte in presence of NHIF enhances the value further.

The room temperature (25 °C) battery performance of the highest conducting 5NHIF0.6LiPy ionanofluid electrolyte is compared with popular 0.2LiPyr$_{14}$TFSI electrolyte by assembling Li/electrolyte +GF/A separator/LMO cells. Fig. 6(a) and 6(c) depict the charge-discharge profile of the 0.2LiPyr$_{14}$TFSI and 5NHIF0.6LiPy cells recorded with various current densities.The comparison of charge-discharge profiles of the 1st, 20th and 50th cycle of the cells at a current density of 0.24 mAg$^{-1}$ are also shown in Fig. 6(b) and 6(d), respectively. For each of the cases, charging profile displays a capacity buildup in the voltage range of 4.0-4.2 V (vs. Li/Li$^+$), while the discharge profile exhibits sloping characteristics, with an average discharge voltage of approximately 3.8 V, typical for Li/LiMO cell [29]. The measured discharge capacity of LMO, using the 5NHIF0.6LiPy electrolyte at a current density of 0.24 mAg$^{-1}$ is 131 mAhg$^{-1}$, which is 88.5% of the theoretical value (148 mAh g$^{-1}$) of LMO [29] and significantly higher than the value (87 mAhg$^{-1}$, 59% ) obtained in 0.2LiPyr$_{14}$TFSI electrolyte. This may be attributed to the observed difference in the $t_{Li}{}^{+}$ value of the electrolytes. The difference in battery performance between the cells become more significant at higher current densities. While the discharge capacity in the ionanofluid electrolyte is found to be 82 mAhg$^{-1}$ at 96 mAg$^{-1}$ current density, 0.2LiPyr$_{14}$TFSI provide only 51 mAhg$^{-1}$ at the similar condition. Moreover, as compared to pure IL-based electrolyte cell, the ionanofluid electrolyte cell demonstrates lesser capacitance fading after long-term reversible cycling operation. This is attributed to the better solid electrolyte interface (SEI) formation capability of the ionanofluid electrolyte in the cell, which is further clarified by the rate performances of the LMO cathode with the electrolytes (Fig. 7(a)). The discharge capacity value (87 mAhg$^{-1}$) obtained in 0.2LiPyr$_{14}$TFSI and in 5NHIF0.6LiPy (131 mAhg$^{-1}$) for the slow rate of 24mAg$^{-1}$ (cycle 1-10) fall up to 51 mAhg$^{-1}$ and 80 mAhg$^{-1}$, respectively at moderately higher current density of 96 mAg$^{-1}$ (cycle 31-40). To check the capacity restoration capability the cells are further cycled at 24mAg$^{-1}$ for additional 10 cycles. Both the cells show >98 % capacity restoration capability.

Finally, the discharge capacity and Coulombic efficiency at a low current density (24 mAg$^{-1}$) of different cells with cycle number is plotted in Fig. 7(b).

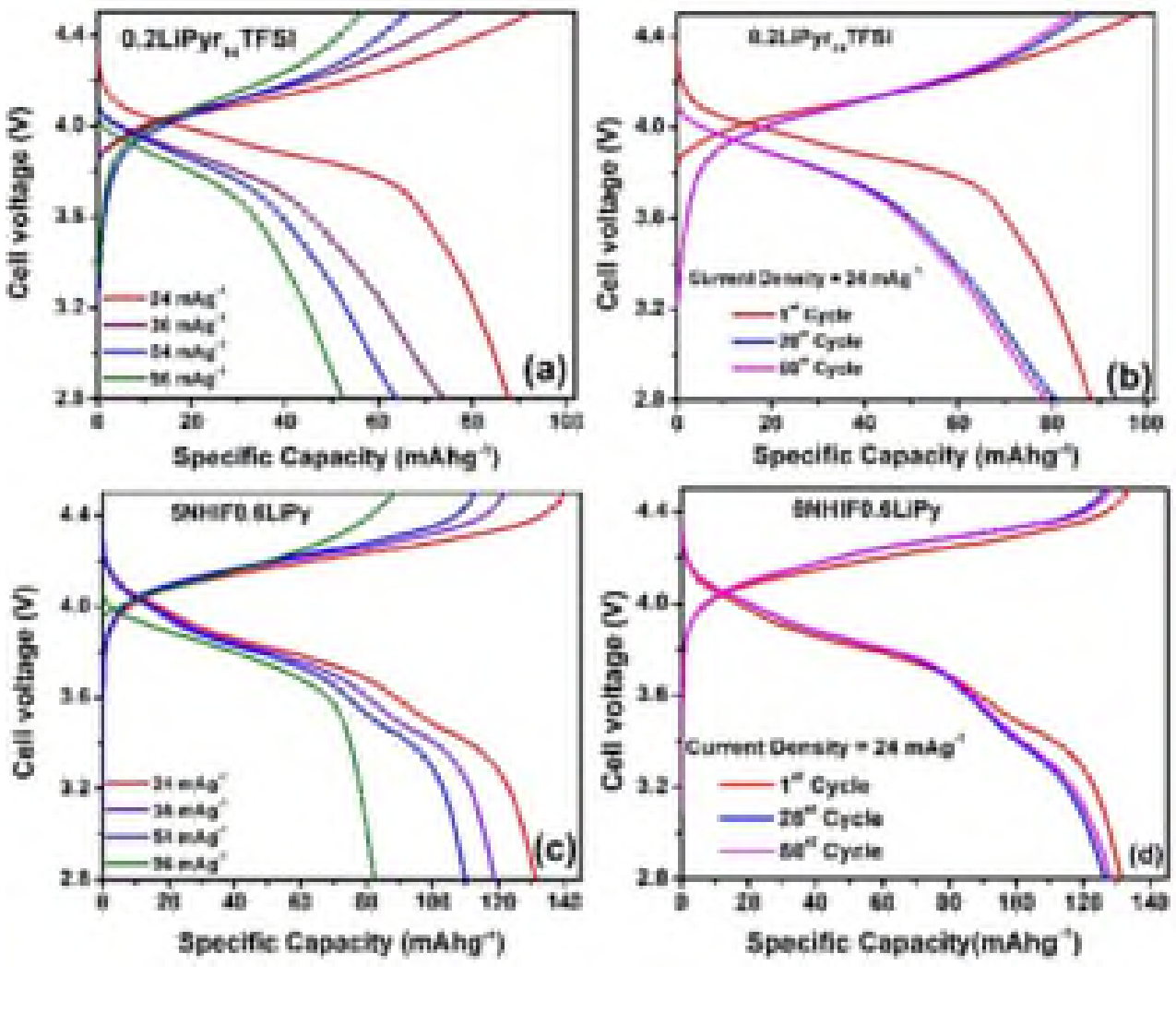

Fig. 6 Charge-discharge profiles at different current densities and at three different cycles for (a)-(b) 0.2LiPyr$_{14}$TFSI and (c)-(d) 5NHIF0.6LiPyr electrolyte comprising Li/$LiMn_2O_4$ cells at 25 °C.

For ionanofluid electrolyte cell, discharge capacity slightly decreases during initial 10-15 cycles and after the formation of stable SEI, capacity improves and stabilizes for the rest of the cycles and retains 96% of the capacitance value after 50 cycles with a Columbic efficiency of 90%. On the other hand, for 0.2LiPyr$_{14}$TFSI electrolyte cell, the initial steady decrease in discharge capacity continues further and finally stabilizes and retain 88% of its initial capacitance with a Columbic efficiency of only 78%.

Thus, both the polarization effect and increase in interfacial resistance during long-term cycling are minimized in presence NHIF in the ionanofluid electrolyte to provide improved rate capability and cyclic ability with the better Columbic efficiency of the cell.

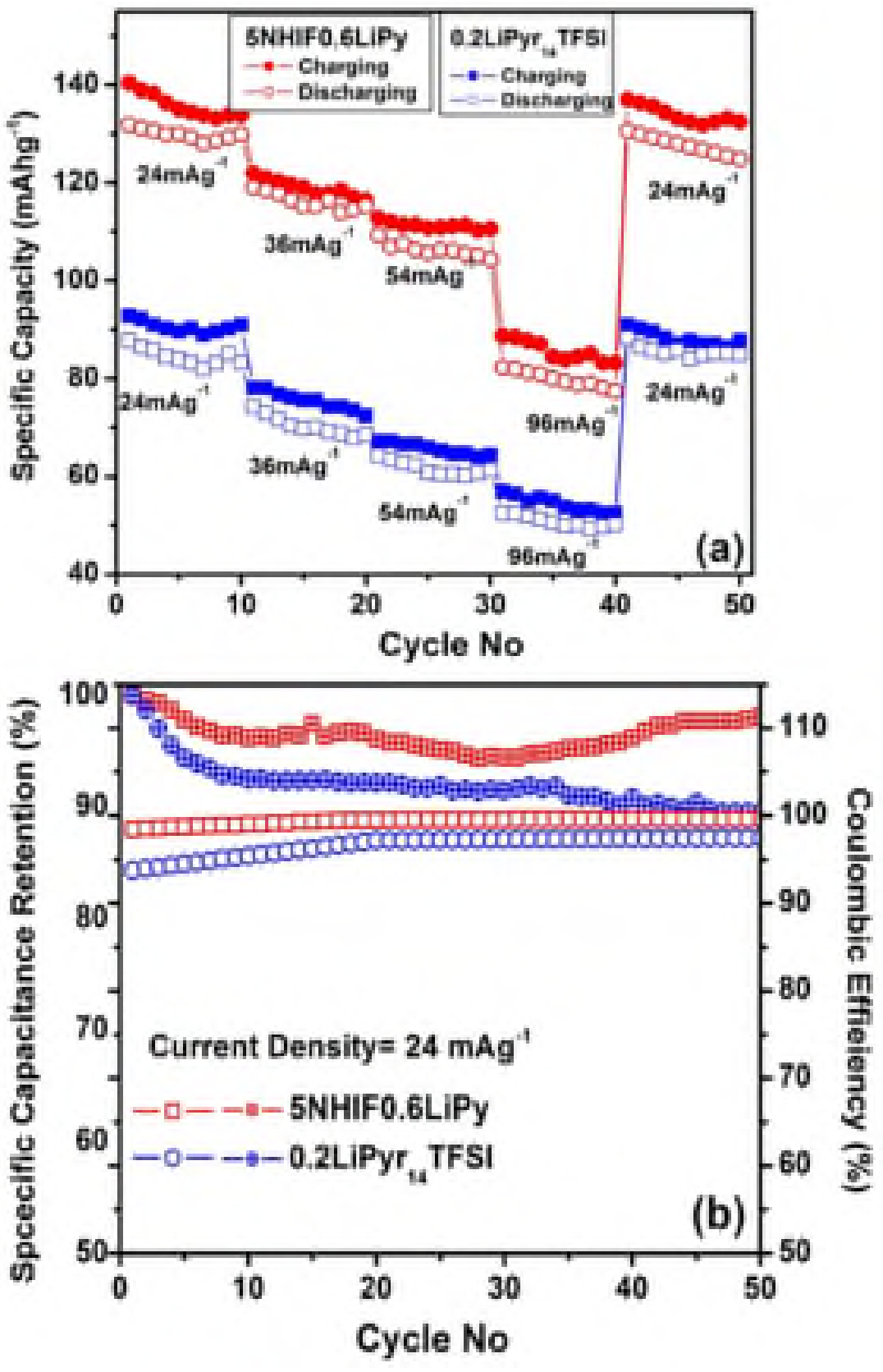

**Fig. 7** (a) Rate performance of the LMO cathode with different electrolytes. (b) Cycling stability and Columbic efficiency of the Li/LMO cell with different electrolytes at the current density of 0.24 mA g$^{-1}$.

## Conclusions

In summary, a unique class of ionanofluid electrolytes have been presented for the application in LMBs. The electrolytes are formed by dispersing ionic liquid functionalized TiO$_2$ nanoparticles in a higher amount of lithium salt dissolved Pyr$_{14}$TFSI electrolyte. Apart from the advantageous properties of IL, the electrolytes exhibit several remarkable features, among which the viscosity decoupled ionic conductivity and suppression of crystallinity in presence of higher lithium salts, are favourable for LMB applications. The electrolytes possess higher ionic conductivity, as compared to much lower lithium salt dissolved Pyr$_{14}$TFSI IL-based electrolyte, along with substantially higher lithium ion transference number at ambient temperature. At 25 °C, the Li/LiMn$_2$O$_4$ cell containing the optimized highest conducting electrolyte deliver more than 88% of the theoretical discharge capacity at a current density of 24 mAg$^{-1}$, much higher than that obtained (59%) from conventional pure IL-based electrolyte cell. Moreover, the ionanofluid electrolyte shows noticeably superior performance in terms of cycling ability and rate capability, as compared to the pure IL-based electrolyte.

### Conflicts of interest

There are no conflicts to declare.

### Acknowledgements

Authors thankfully acknowledge the DST-SERB (Govt. of India) for the financial support under the Research Scheme No: **EMR/2014/000290** Dated 11.09.2015. In addition, the authors thankfully acknowledge the DST-FIST scheme of the Department of Physics, University of Kalyani for providing the instrumental facilities.

# Supporting Information

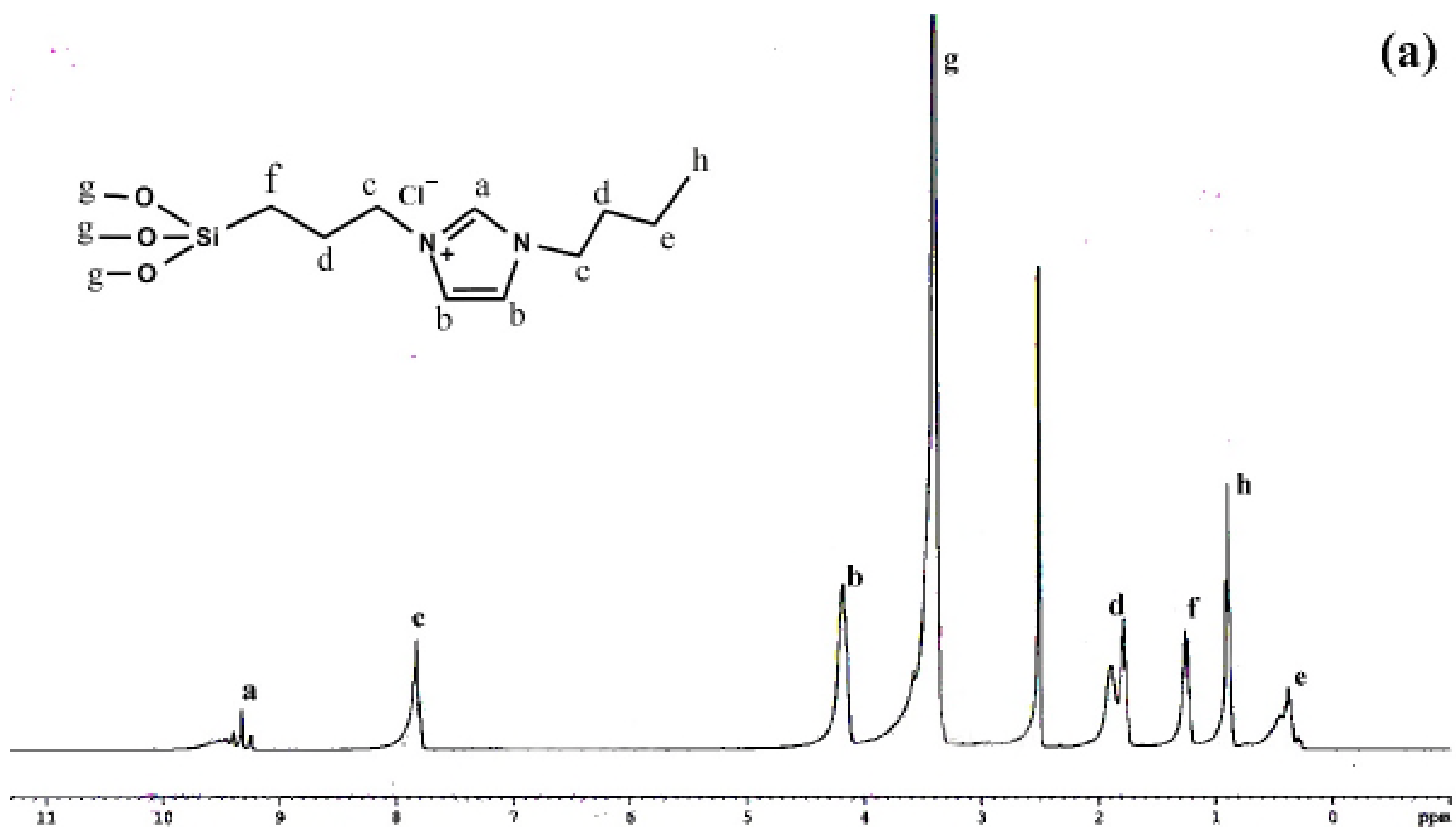

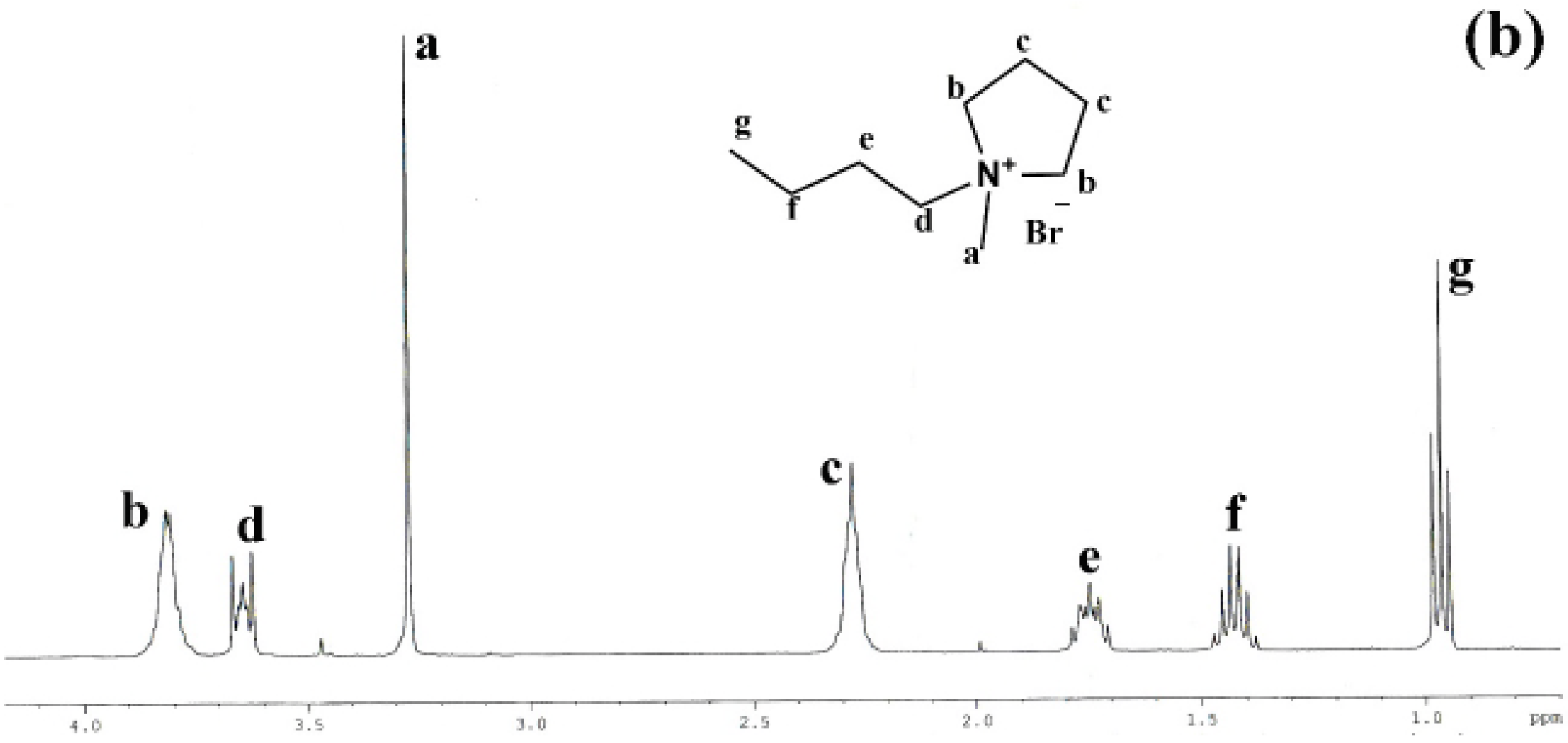

**Fig. S1** $^{1}H$ NMR spectrum of as synthesized (a) 1-*n*-butyl-(3-(trimethoxysilylpropyl)imidazolium chloride and (b) $Pyr_{14}TFSI$ ionic liquids.

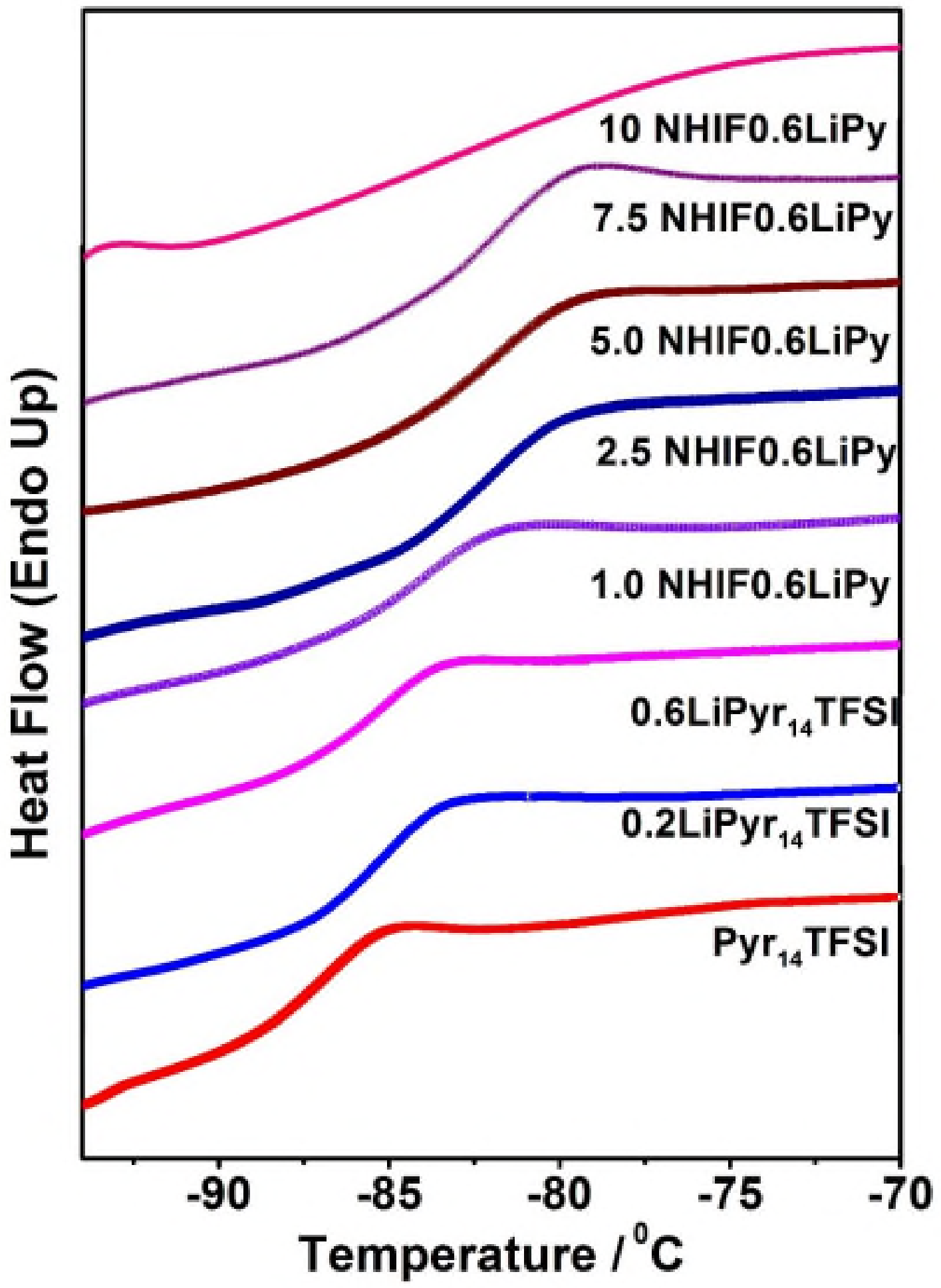

**Fig. S2** Compositional variation of glass transition temperature ($T_g$).

**Table S1** Comparison of thermal characteristics of different electrolytes with $Pyr_{14}TFSI$ ionic liquid during heating @5 °C/min from the respective supercooled states.

| Samples | $T_g$ (°C) | $T_c$ (°C) | | $T_m$ (°C) | | $\Delta H_c$ (J/g) | | $\Delta H_m$ (J/g) | |
|---|---|---|---|---|---|---|---|---|---|
| | | $T_{c1}$ | $T_{c2}$ | $T_{m1}$ | $T_{m2}$ | $\Delta H_{C1}$ | $\Delta H_{C2}$ | $\Delta H_{m1}$ | $\Delta H_{m2}$ |
| Pyr14TFSI | -86.5 | -54.1 | | -27.7 | -17.8 | 48.2 | | 14.2 | 25.2 |
| 0.2LiPyr14TFSI | -85.4 | -32.5 | | -22.7 | | 13.1 | | 14.2 | |
| 0.6LiPyr14TFSI | -84.5 | -53.8 | -23.4 | -6.6 | | 22.8 | 14.8 | 45.6 | |
| 1NHIF0.6LiPy | -83.1 | -53.9 | | -17.72 | | 23.3 | | 25.4 | |
| 2.5NHIF0.6LiPy | -82.2 | -53.3 | | -17.8 | | 21.2 | | 24.4 | |
| 5NHIF0.6LiPy | -81.3 | -51.7 | | -18.1 | | 14.3 | | 17.1 | |
| 7.5NHIF0.6LiPy | -79.3 | -27.7 | | -18.3 | | 12.5 | | 13.6 | |
| 10NHIF0.6LiPy | -77.5 | | | | | | | | |

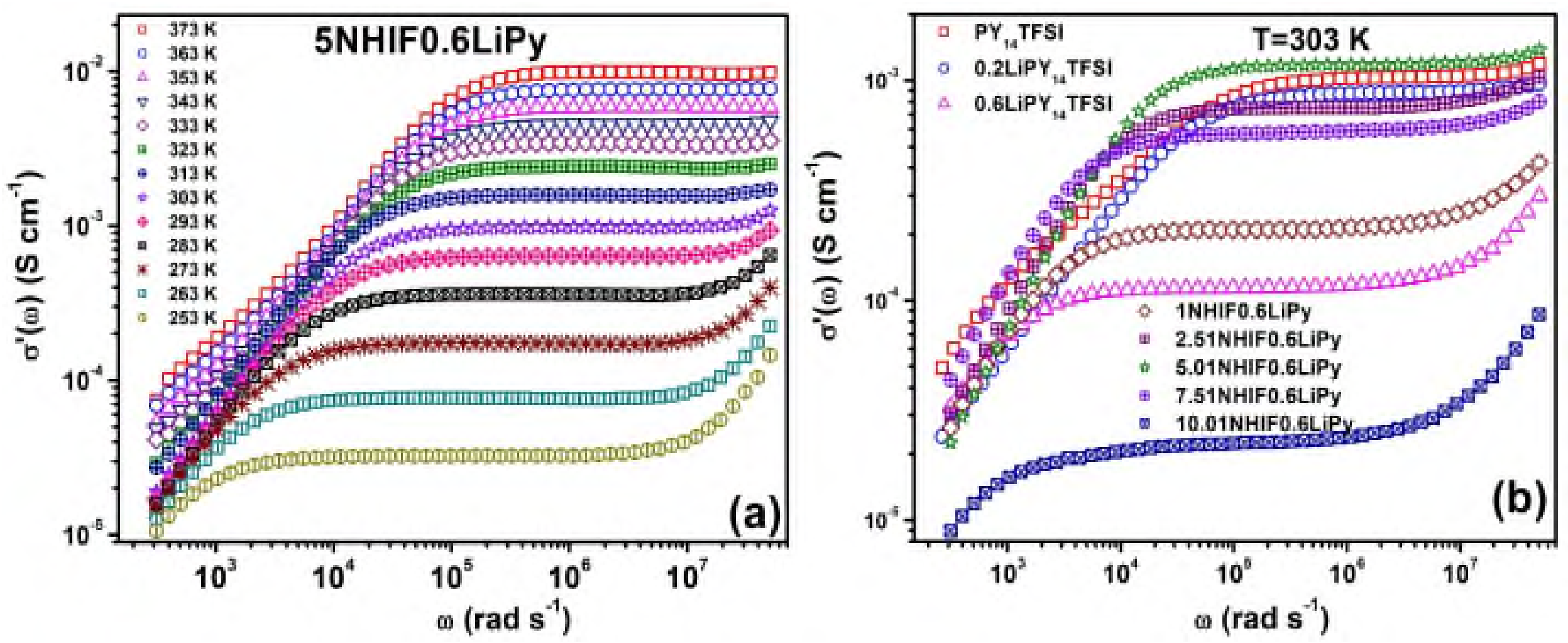

**Fig. S3** Real part of complex conductivity spectra as a function of (a) temperature for a representative sample and (b) composition at a particular temperature.